# EFFICIENT ECC-BASED AUTHENTICATION SCHEME FOR FOG-BASED IOT ENVIRONMENT

Mohamed Ali Shaaban[1], Almohammady S. Alsharkawy[2], Mohammad T. Abou-Kreisha[2] and Mohammed Abdel Razek[2]

[1] Department of Basic Science, Faculty of Engineering, Sinai University, Egypt
[2] Department of Mathematics, Faculty of Science, Al-Azhar University, Nasr City, Egypt

*ABSTRACT*

*The rapid growth of cloud computing and Internet of Things (IoT) applications faces several threats, such as latency, security, network failure, and performance. These issues are solved with the development of fog computing, which brings storage and computation closer to IoT-devices. However, there are several challenges faced by security designers, engineers, and researchers to secure this environment. To ensure the confidentiality of data that passes between the connected devices, digital signature protocols have been applied to the authentication of identities and messages. However, in the traditional method, a user's private key is directly stored on IoTs, so the private key may be disclosed under various malicious attacks. Furthermore, these methods require a lot of energy, which drains the resources of IoT-devices. A signature scheme based on the elliptic curve digital signature algorithm (ECDSA) is proposed in this paper to improve the security of the private key and the time taken for key-pair generation. ECDSA security is based on the intractability of the Elliptic Curve Discrete Logarithm Problem (ECDLP), which allows one to use much smaller groups. Smaller group sizes directly translate into shorter signatures, which is a crucial feature in settings where communication bandwidth is limited, or data transfer consumes a large amount of energy. In this paper, we have chosen the safe curve types of elliptic-curve cryptography (ECC) such as M-221, SECP256r1, curve 25519, Brainpool P256t1, and M-551. These types of curves are the most secure curves of other curves of ECC as their security is based on the complexity of the ECDLP of the curve. And these types of curves exceed the complexity of the ECDLP. A valid signature can be generated without re-establishing the whole private key. ECDSA ensures data security and successfully reduces intermediate attacks. The efficiency and effectiveness of ECDSA in the IoT environment are validated by experimental evaluation and comparison analysis. The results indicate that, in comparison to the two-party ECDSA and RSA, the proposed ECDSA decreases computation time by 65% and 87%, respectively. Additionally, as compared to two-party ECDSA and RSA, respectively, it reduces energy consumption by 77% and 82%.*

## 1. INTRODUCTION

The advantages of a huge data warehouse, the ability to perform complicated computations, and the accessibility of data from anywhere in the world are all characteristics of cloud computing. Despite this, the centralization idea of the cloud computing paradigm requires all processes to take place in the same place [1]. Cloud computing is unsuitable for applications that need high mobility, location identification, and low latency due to its centralised structure. Fog computing was therefore developed as a new concept to overcome the drawbacks of cloud computing [2]. Fog computing is a new architecture that serves as a layer between the Internet of Things (IoT) and cloud worlds in order to bring services to the network edge.

           

Fog computing will improve service quality for reporting live applications, reduce latency, and improve location sensitivity [3]. Additionally, it offers a potentially efficient way to manage the massive amount of data generated by users, hardware, and software. The fog is a cloud-alternative technology that aims to satisfy users' requirements for security, latency, efficiency, cognition, and agility [4]. Fog computing distributes resources to nearby virtual clusters, whereas cloud computing aims to reduce resources globally. This is the primary distinction between the two types of computing [5]. Additionally, the fog differs in terms of mobility assistance, geographical coverage, and proximity to the end user.

In the fog computing architecture, the edge devices such as vehicles, sensors, actuators, or user information software are the most fundamental level of this architecture. Fog nodes are located in the second tier and are used to store, process, or transmit data to the cloud from edge devices. Fog nodes use an excessive number of protocols, like Wi-Fi or Bluetooth, to collect data from edge devices. The top layer is the cloud data center, from which fog nodes send data [6]. Edge layer is the closest to the user and the real world. It includes many IoT subsets including smartphones, sensors, smart cards, and intelligent cars etc. Typically, these sensors are scattered extensively and detect and collect the feature information of actual events or objects before transferring the perceived information to the layers above, either for saving or processing. Fog layer is located at the edge of the network, this layer is made up of several fog nodes, including switchers, routers, access points, gateways, fog servers, base stations, etc. These fog nodes are widely scattered between the cloud and end devices, including places like malls, bus stops, cafes, parks, streets, etc. The fog nodes enable connectivity with end devices to supply services regardless of their position, whether they are moving on trucks or stationary at a site. They are capable of transferring, quantifying, and storing data obtained through sensing. Actual-time analysis and latency-sensitive operations are performed at the fog tier. Additionally, by connecting to the cloud's data center via the IP core network, fog nodes can collaborate and communicate with the cloud to obtain more advanced strengths for saving and processing. Cloud layer, which is mostly made up of high-powered servers and storage niches, is in charge of providing a variety of services for applications including smart factories, smart homes, smart offices, and smart transportation. This tier can store, save, and compute with enormous power; thus, it can perform complex computer analysis and store and preserve enormous amounts of data and information.

There are various major cloud computing security dangers that incorporate critical security measures to assure in the cloud. Nonetheless, measurements are unsuitable for fog computing due to their various mobility, characteristics, large-scale geo-dispersion, and heterogeneity. Furthermore, the fog is an appealing target for digital attackers because it contains enormous amounts of sensitive data from both IoT and cloud devices. In this way, further research is needed to increase security. the Security and Privacy threats that faced in Fog Computing such as Security of networks, Authentication, Privacy of Identity, Privacy of Location Data, Problem of Malicious Fog Node, Fog Computing Environments-Malicious detection techniques, and attacks of Replay [7].

Nowadays, IoT connects with applications that require high-speed data processing, location awareness, and low energy consumption while managing massive amounts of generated data. IoT-devices need the capacity to handle all these tasks; thus, they are related to fog computing services. [8].

To preserve data security in a fog computing environment and to safeguard fog users' privacy, a variety of security services must be met. Among these services, anonymity, authentication, replay attack, impersonation attack, perfect forward secrecy, confidential communication, session keys independence, fog user/server compromise, and man-in-the-middle attack are the most crucial. [9, 10].

Authentication is the most important security factor since without it, a hacker might pretend to be a fog user or server. consequently, the end user must authenticate within the fog network to use fog services. On the other hand, due to the disparate power, storage, and processing capabilities of the end devices, the authentication process in the fog network is all but impossible. In addition to the assortment of end-user capabilities, they continually have a limited amount of resources. The researchers are consequently unable to implement the authentication process using widely used methods, such as public key infrastructure (PKI) [11].

This article discusses the effective fog authentication protocol. Hence, in this paper, ECDSA has been proposed to improve data security and privacy in data transmission. To authenticate users, the literature suggests various approaches that differ in computing cost, speed, and threat. Nonetheless, developing lightweight algorithms is difficult because of the resource constraints of IoT-devices [12, 13]. What makes ECDSA attractive is that its security is based on the intractability of the Elliptic Curve Discrete Logarithm Problem (ECDLP), which allows one to use much smaller groups compared to its classical counterpart, RSA, and two-party ECDSA. Smaller group sizes directly translate into shorter signatures, which is a crucial feature in settings where communication bandwidth is limited, or data transfer consumes a large amount of energy. RSA security rests on the integer factorization problem (IFP). The RSA algorithm's efficiency suffers significantly due to its large key size, making it unsuitable for local systems. Two-party ECDSA protocol used artificial and interactive assumptions on the Paillier scheme, proving the security of the protocol under malicious party two. Uses such as guessing, aborting a protocol under malicious two, and sending encrypted secret shares with Paillier require additional, high-cost checks. A lightweight cryptography solution is also required to encrypt such data in privacy-aware applications due to the high data volume produced by IoT-devices. Therefore, this paper presents a lightweight ECC-based authentication scheme for IoT-based cloud environments. We will use safe ECC curves such as M-221, SECP256r1, curve 25519, Brainpool P256t1, and M-551 to ensure communication between cloud servers and IoT-devices.

The proposed scheme using ECDSA for digital signature has some implications in IoT-Fog environments. The IoT-Fog environment will benefit from using the proposed authentication scheme for digital signatures since it will conserve network resources, reduce processing time, and reduce energy consumption while maintaining the highest levels of security and preserving the privacy of the users.

The article's significant contributions are as follows:
- We investigate the issue of data authentication in an IoT environment.
- We provide an efficient authentication protocol that offers device authentication for the IoT-based Fog environment.
- We evaluated the effectiveness of five curve types that are used with our protocol, ECDSA, to authenticate IoT-devices. Performance is evaluated considering energy consumption and computation time.

The rest of the paper is organized as follows: An overview of the related works is provided in Section 2. Section 3 preliminaries the methods used. Section 4 illustrates the proposed methodology and the integration of ECC and application. Our experimental evaluation approach is described in Section 5, along with a discussion of the outcomes. The paper is concluded in section 6.

## 2. RELATED WORKS

Several security concerns, such as worldwide cyber threats and privacy protection, are brought on by the Internet of Things. Researchers have proposed reliable and effective authentication strategies to address security challenges in the literature on lightweight authentication mechanisms. Fog devices are subject to different risks at the network edge that are handled by well-managed cloud systems. The most important step in securing an IoT application in a fog computing environment is authentication.

To date, many two-party digital signatures have been proposed. The Security of Practical Two-Party RSA Signature Schemes proposed by Bellare and Sandhu [14] suggested two notions of security for two-party signature schemes and provided proofs of security for the schemes based on assumptions about RSA and the hash function. The RSA public key algorithm was developed by Ron Rivest, Adi Shamir, and Leonard Adleman in 1977. Soon, it became popular and was adopted by many standard bodies. It is based on the difficulty of finding the factors for the product of two large prime numbers. Key generation in the RSA digital signature scheme is the same as key generation in the RSA cryptosystem. RSA signature generation and validation RSA encryption compute s, where s is the signature and RSA decryption compute m using signature (s). Signature validation consists of validating m, which has computed in RSA decryption. The RSA digital signature scheme takes a long time, which is not appropriate for IoT-devices. A Two-Party Generation of DSA Signatures proposed by MacKenzie and Reiter [15] Until 2017 it was the most efficient DSA/ECDSA two-party threshold signature scheme. It was based on multiplicative secret shares of ECDSA private key $d$ and instance key $k$, it uses Paillier encryption and ZKP (zero-knowledge proof) to protect from malicious adversaries. Such ZKP causes this scheme to work inefficiently: digital signature generation takes several seconds, which is not appropriate for modern communication. and the security of the algorithm was proved with random oracle model. Zhang [16] proposed a practical distributed two-party SM2 signature algorithm. SM2 algorithm is issued by the Chinese Government's State Cryptography Administration.

A two-party ECDSA protocol proposed by Lindell [17], which is faster than previous protocols. In this paper, Lindell used artificial and interactive assumption on Paillier scheme proving security of protocol under malicious party $P2$. In other terms, uses guessing aborting a protocol under malicious $P2$ .and sending encrypted secret share $d1$ with Paillier requires additional highly cost range checks: $d1$ must be both in message space of Paillier cryptosystem and less than n (an order of cyclic subgroup generated by point G). D. He and Y. Zhang [18] proposed two-part identity-based signature protocol in 2018. However, identity-based cryptography has a problem with key escrow. Zhong. et al. [19] have presented an effective ECDSA for devices with limited resources, such as sensor networks. In [20], they presented an ECDSA method without the need for inversion. Their job, however, entails additional elliptic curve multiplication operations, which reduces their efficiency. Jan. et al. [21] have provided an effective mutual authentication system for the servers and the IoT-devices. This system uses a shared essential technique for authentication, and to share a secret between two parties, the DTLS protocol is used. As a result, the suggested technique requires specific security enhancements. Wang. et al. [22] have presented an Effective and secure data transmission that requires client-aware negotiation. As a method of authentication for IoT-devices, they implemented ECDHE-RSA. However, the RSA algorithm's efficiency suffers significantly due to its large key size, making it unsuitable for local systems. The authors in [23] have proposed a solution for initialization and providing the OTP based on the Mobile Agent (IPOM). also proposed a security routing protocol using a one-time password authentication mechanism based on mobile agents (AODVMO) by extending the original AODV protocol and integrating the IPOM solution. Analysis results confirm that AODVMO can prevent almost all current routing protocol attack types, such as blackhole, sinkhole, grayhole, whirlwind, and wormhole types.

Based on a thorough study of existing systems, it is challenging to develop a secure authentication and resource-efficient technique that deal with the limitations of the IoT-Fog environment. Some proposed methods needed to be sufficiently secure, while others were too large and complex for devices with limited resources. We present a secure, lightweight authentication method using the ECC and ECDSA to address these issues and enable device authentication for the IoT-Fog environment.

## 3. PRELIMINARIES

There are many significant cloud computing security dangers that incorporate critical security measures to ensure security in the cloud. Nonetheless, measurements are unsuitable for fog computing due to their various mobility, large-scale geo-dispersion, and heterogeneity. Furthermore, the authors in [24, 25] illustrated that the potential security issues of an IoT-fog environment include security of the network, authentication, privacy of identity, the problem of malicious fog nodes, and forgery. To some extent, fog computing engineering can overcome the security challenges of IoT cloud design.

The ECC is a key-based technique for encrypting data that is based on the ECDLP's difficulty and the algebraic structures of elliptic curves over finite fields. ECC is increasingly used in the implementation of public-key cryptography protocols. ECC implements encryption, key exchange, and signatures, all fundamental features of asymmetric cryptosystems. The effectiveness of asymmetric cryptography is based on the supposed difficulty of problems in number theory, such as clusters of points on an elliptic curve, the computation of discrete logarithms in finite fields, and integer factorization.

ECC offers many advantages, especially in limited resource environments, and provides the same level of security as other frequently used methods like RSA, but with smaller key sizes and more effective implementations. For instance, ECC offers substantially shorter keys, up to 160 bits instead of the 1024 bits used by RSA. ECC is an effective system to use in IoT-fog environments because it requires less processing, less power, less storage, less memory, and less bandwidth than other systems. Therefore, ECC offers a higher level of security while using less computation and bandwidth. The ECC-based authentication schema has more security challenges such as Replay attack, Forgery attack, Impersonation attack and Man-in-the-Middle attack. The proposed protocol resists all these attacks. However, because of elliptic curve cryptography's (ECC) outstanding performance and low-key size specifications, the development of an effective IoT authentication protocol based on ECC is achievable. ECC has already been utilized in several authentication and key establishment protocols for IoT-based devices. ECC is a public key or asymmetric key cryptography solution that provides smaller key sizes and lower computation costs than alternatives such as RSA or systems that rely on the discrete logarithm as a trapdoor.

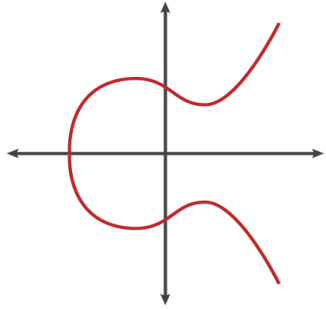

Figure 1. Elliptic Curve graph.

The elliptic curve E(Q) over a field Q is shown in Figure 1. The E(Q) is an algebraic cubic smooth projective curve with base point φ that is described, and the points on E(Q) form an algebraic group with identity point φ. Any elliptic curve is isomorphic to a cubic curve of form Eq.1 according to the Riemann-Roch theorem. [26–27].

$$E(Q) = \{(x,y) \mid y^2 + b_1xy + b_3y = x^3 + b_2x^2 + b_4x + b_6\} \cup \{\varphi\} \quad (1)$$

Where b1, b2, b3, b4, and b6 are Q-rational coefficients that satisfy a non-zero discriminant condition and φ is the point at infinity. Elliptic curves are represented in several forms for computational objectives in cryptography as follows:

1. Weierstrass curves: An elliptic curve over GF(q) for some prime q constitutes a short Weierstrass curve with form Eq.2:
$$E(GF(q)) = \{(x,y) \mid y^2 = x^3 + ux^2 + v\} \cup \{\varphi\} \quad (2)$$
where u and v are Q-rational coefficients such that 4u3+27v2 is non-zero. The most common elliptic curve representations are Weierstrass curves since they are isomorphic to all elliptic curves over fields with (characteristic > 3).

2. Montgomery curves: is elliptic curve over GF(q) for some prime q, with the form Eq.3:
$$E(GF(q)) = \{(x,y) \mid vy^2 = x^3 + ux^2 + x\} \cup \{\varphi\} \quad (3)$$
where v and u are Q-rational coefficients whereas (v(u2 - 4) ≠ 0). When performing calculations, Montgomery curves only use the first affine coordinate and can efficiently multiply using the Montgomery ladder.

3. Edwards curves: is an elliptic curve over GF(q) for some prime q, with the form Eq.4:
$$E(GF(q)) = \{(x,y) \mid ux^2 + y^2 = 1 + vx^2y^2\} \quad (4)$$
where v and u are Q-rational coefficients whereas (v(1-v) ≠ 0). For Edwards curves, the addition and doubling formulas converge and there is no point at infinity.

### 3.1. The ECC Implementation

As indicated in Figure 2, an ECC hierarchy has four successive levels. Finite field arithmetic operations are included in the first level and can be implemented in Galois binary field GF($2^m$) and Galois prime field GF(p). Operations on E(Q) groups, like point addition (PA) and point doubling (PD), are included at the second level. At the third level, elliptic curve group operations are successively combined to produce elliptic curve point multiplication (ECPM). ECC protocols like ECDSA and ECDH (Elliptic curve Diffie-Hellman) are included at the highest level. The most critical process in an ECC system is ECPM.

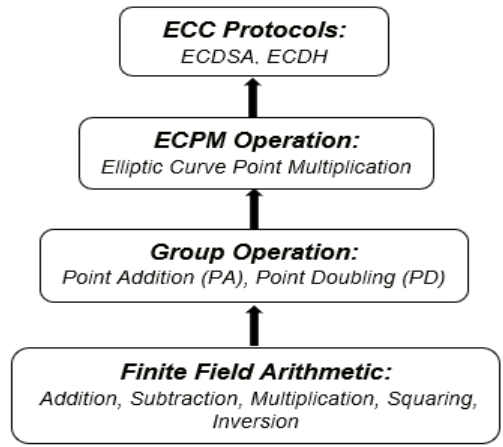

Figure 2. The ECC implementation hierarchy.

## 4. THE PROPOSED METHODOLOGY

The major components and architecture of the proposed ECC-based Authentication for protecting IoT in the fog layer are presented in this section. This section's objective is to review the physical and functional components of the proposed approach and how they contribute to the system's operational concept.

Various standards indicate that numerous elliptic curves can be used in ECC. Prime field size and curve form are the main factors utilized to describe several curve types. In the proposed scheme, we selected five types of ECC, which are highly prevalent in ECC applications, to apply in IoT-Fog environment.

1. **M-221 Curve**: M-221 is 221-bit prime field Montgomery curve, it has equation as $By^2 = x^3+Ax^2+x$, and Table 1 contains the domain parameters of M-221 curve [28].
2. **M-511 Curve**: M-511 is 511-bit prime field Montgomery curve, it has equation as $By^2 = x^3+Ax^2+x$, and Table 1 contains the domain parameters of M-511 curve [29].
3. **Brainpool P256t1 Curve**: Brainpool P256t1 is 256-bit prime field Weierstrass curve, it has equation as $y^2 = x^3+ax+b$, and Table 1 contains the domain parameters of Brainpool P256t1 curve [30].
4. **SECP256r1 Curve**: SECP256r1 is 256-bit prime field Weierstrass curve, it has equation as $y^2 = x^3+ax+b$, and Table 1 contains the domain parameters of SECP256r1 curve [31].
5. **25519 Curve**: Curve 25519 is 255-bit prime field Montgomery curve, it has equation as $By^2 = x^3+Ax^2+x$, and Table 1 contains the domain parameters of 25519 curve [32].

Each curve has distinct field widths, which are the ECC key sizes. The chosen curves are then analysed using the ECC techniques called ECDSA. For each curve, the same procedure is utilized to do the analysis. Only the curve parameters vary depending on the kind of curve. The computation time required for all ECDSA operations is calculated for each curve type.

| M-221 Curve Parameters | |
|---|---|
| p | 3369993333393829974333376885877453834204643052817571560137951281149 |
| a | 01c93a |
| b | 01 |
| G | (04, 0f7acdd2a4939571d1cef14eca37c228e61dbff10707dc6c08c5056d) |
| n | 421249166674228746791672110734682167926895081980396304944335052891 |
| h | 8 |
| **M-511 Curve Parameters.** | |
| p | 67039039649712985497870124991029230637396829102961966888617807218608820150367734884009371490834517138450159290932430254268769414059732849732168245030041861 |
| a | 81806 |
| b | 01 |
| G | (05,2fbdc0ad8530803d28fdbad354bb488d32399ac1cf8f6e01ee3f96389b90c809422b9 429e8a43dbf49308ac4455940abe9f1dbca542093a895e30a64af056fa5) |
| n | 8379879956214123187233765623878653829674603637870245861077225902326102518796074108048767793830555087621410592584974489349870525087756261624609307379422 99 |
| h | 8 |
| **Brainpool P256t1Curve Parameters** | |
| p | 76884956397045344220809746629001649093037950200943055203735601445031516197751 |
| a | a9fb57dba1eea9bc3e660a909d838d726e3bf623d52620282013481d1f6e5374 |
| b | 662c61c430d84ea4fe66a7733d0b76b7bf93ebc4af2f49256ae58101fee92b04 |
| G | (a3e8eb3cc1cfe7b77732213b23a656149afa142c47aafbc2b79a191562e1305f4, 2d996c823439c56d7f7b22e14644417e69bcb6de39d027001dabe8f35b25c9be) |
| n | 76884956397045344220809746629001649092737531784414529538755519063063536359079 |
| h | 1 |
| **SECP256r1 Curve Parameters** | |
| p | 115792089210356248762697446949407573530086143415290314195533631308867097853951 |
| a | -3 |
| b | 5ac635d8aa3a93e7b3ebbd55769886bc651d06b0cc53b0f63bce3c3e27d2604b |
| G | (6b17d1f2e12c4247f8bce6e563a440f277037d812deb33a0f4a13945d898c296, 4fe342e2fe1a7f9b8ee7eb4a7c0f9e162bce33576b315ececbb6406837bf51f5) |
| n | 115792089210356248762697446949407573529996955224135760342422259061068512044369 |
| h | 1 |
| **25519 Curve Parameters** | |
| p | 57896044618658097711785492504343953926634992332820282 19728792003956564819949 |
| a | 76d06 |
| b | 01 |
| G | (09, 20ae19a1b8a086b4e01edd2c7748d14c923d4d7e6d7c61b229e9c5a27eced3d9) |
| n | 7237005577332262213973186563042994240857116359379907606001950938285454250989 |
| h | 8 |

The security and efficiency of ECC based on the complexity of the ECDLP of the curve. The authors in [33], evaluated each curve for its ECC security and ECDLP. Four parameters are considered for ECDLP security as follows:
- Curve security against Pollard's Rho algorithm.
- The security against multiplicative transfers.
- The curve rigidity.
- The complex multiplication field discriminator.

If any of the above-mentioned parameters are not found for the curve, the curve is regarded as unsecure for use in the applications of ECC. The ECDLP security parameters determine the security of the curve. According to [33], the M-221, Curve 25519, Brainpool P256t1, SECP256r1, and M-511 curves are secure curves as they satisfy the ECDLP security parameters' rigidity requirement.

### 4.1. The ECDSA protocol

The ECDSA is the DSA's elliptic curve analogue. this protocol requires integer multiplication, inverse operation, modular operation, and a hash function in addition to elliptic curve operations like scalar field multiplication, and field inverse multiplication. In the ECDSA, p1 (person) creates the signature with his private key, and p2 (another person) verifies the signature with p1's public key. Using ECDSA for digital signatures brings a number of essential advantages, such as:

- Compared to prior cryptography tools such as RSA and DSA, elliptic curve algorithms provide more security for a given key size. This also occurs with small key sizes, which are, by definition, more vulnerable than larger sizes.
- In the case of ECDSA, the time and memory space required to create and deliver messages are substantially less than in earlier technologies.

ECDSA is based upon some parameters common to all entities participating in a network. These parameters are listed as the following:
- The elliptic curve $E$ defined over $F_P$ prime field.
- The prime field $F_P$ size is described by $p$.
- The $a$ and $b$ coefficients of the elliptic curve $E$ equations.
- The base point $G$ in $E(F_P)$.
- The $n$ order of $G$, typically a prime.
- The $h$ cofactor $= |E(F_P)| / order(n)$

### 4.2. The ECDSA Phases

ECDSA is known for its quick signature generation and verification, providing an effective technique against cyberattacks. Key generation, signature generation, and verification are the phases of the ECDSA process. The explanation of these three phases will be detailed as follows:

#### 4.2.1. ECDSA Key Generation

ECDSA key is linked to a specific set of EC domain settings. The public key is a randomly generated multiple of the base point, whereas the private key is the integer used to generate the multiple points. assume that $p_1$ the signatory for a message $M$. Entity $p_1$ performs the following steps to generate a public and private key, the steps of this phase described as follows:
- Choose an elliptic curve $E$ such that the number of points in $E(F_p)$ is divisible by a large prime $n$.
- Choose a base point, $P$, of order $n$ such that $P \in E(F_P)$.
- Choose a random integer $d$, from $[1, n-1]$.
- Calculate $Q = d*P$.
- Sender $p_1$'s private key is $d$.
- Sender $p_1$'s public key is the combination $(E, P, n, Q)$.

### 4.2.2. ECDSA Signature Generation

The signature generation algorithm is based on the ElGamal signature scheme. It takes the private key of the sender and the message to be sent as input and generates the signature as output.

In these steps, using $p_1$'s private key, $p_1$ generates the signature for *M*. The outputs are the signature (r, s), Where the signature components r and s are integers, and using the following steps as follows:
- Select a random integer *k*, from *[1, n-1]*.
- Determine $(x_1, y_1) = k * P$, where $x_1$ is an integer.
- Determine $r = x_1 \bmod n$ If *r = 0*, then go to the first step.
- Determine $h = H(M)$, where *H* is the *(SHA-1)*
- Determine $s = k^{-1}(h + dr) \bmod n$ If *s = 0*, then go to first step.
- The signature of $p_1$ for message *M* is the integer pair *(r, s)*.

### 4.2.3. ECDSA Signature Verification

The signature verification algorithm takes the message and the signature (r, s) as input, and returns a boolean value representing whether the signature is verified.

In this step, the receiver $p_2$ can verify the authenticity of $p_1$'s signature *(r, s)* for message *M* by performing the following steps as follows:
- Obtain signatory $p_1$'s public key *(E, P, n, Q)*.
- Verify that values *r* and *s* are in *[1, n-1]*.
- Compute $w = s^{-1} \bmod n$.
- Compute $h = H(M)$, based on the secure hash algorithm utilized by $p_1$.
- Compute $u_1 = hw \bmod n$ and $u_2 = rw \bmod n$.
- Compute $u_1 P + u_2 Q = (x_0, y_0)$, and $v = x_0 \bmod n$.
- if *v = r*, then the signature of *M* is verified.

### 4.3. The proposed ECDSA scheme

As the ECC offers a higher level of security with a small key size compared with other public key cryptography and reduces the computation power, we offer a lightweight ECC on ECDSA authentication scheme. The proposed scheme is described in Figure 3 consists of two phases: the initialization phase and the IoT-Devices authentication phase, which are explained as follows:

### 4.3.1. Initialization phase

- The fog node requests Key-pair and signature by sending the request with Fog id to the Cloud.
- The Cloud checks Fog node-id, if it exists, then generates two keys and signature, then sends the private and public keys and signature to the Fog.
- The fog node sends a message authentication request signed by a signature algorithm to the Cloud.
- Cloud checks signature verification: if it is valid, then sends "valid" Else, "not valid" to the Fog node.

## 4.3.2. IoT-Devices Authentication

After the initialization phase, all distributed Fog nodes use the ECDSA algorithms to provide keys and authenticate their corresponding IoT-devices. The steps for an IoT-device to get keys and message authentication are as the following:

- IoT-device registers in the cloud.
- The identity provider in the cloud generates IoT-device-id, then IoT-device-id sent to the IoT-device, and to Fog node for the validation part.
- The IoT-device transmits keys and signature request to the Fog node, along with its id.
- The Fog node checks device-id, if it exists, then generates and sends the public key, private key, and signature to IoT-devices.
- The devices send a message authentication request signed by the signature algorithm to the Fog node, then the Fog node checks the signature verification to see if it is valid, then sends validation to the IoT-device.

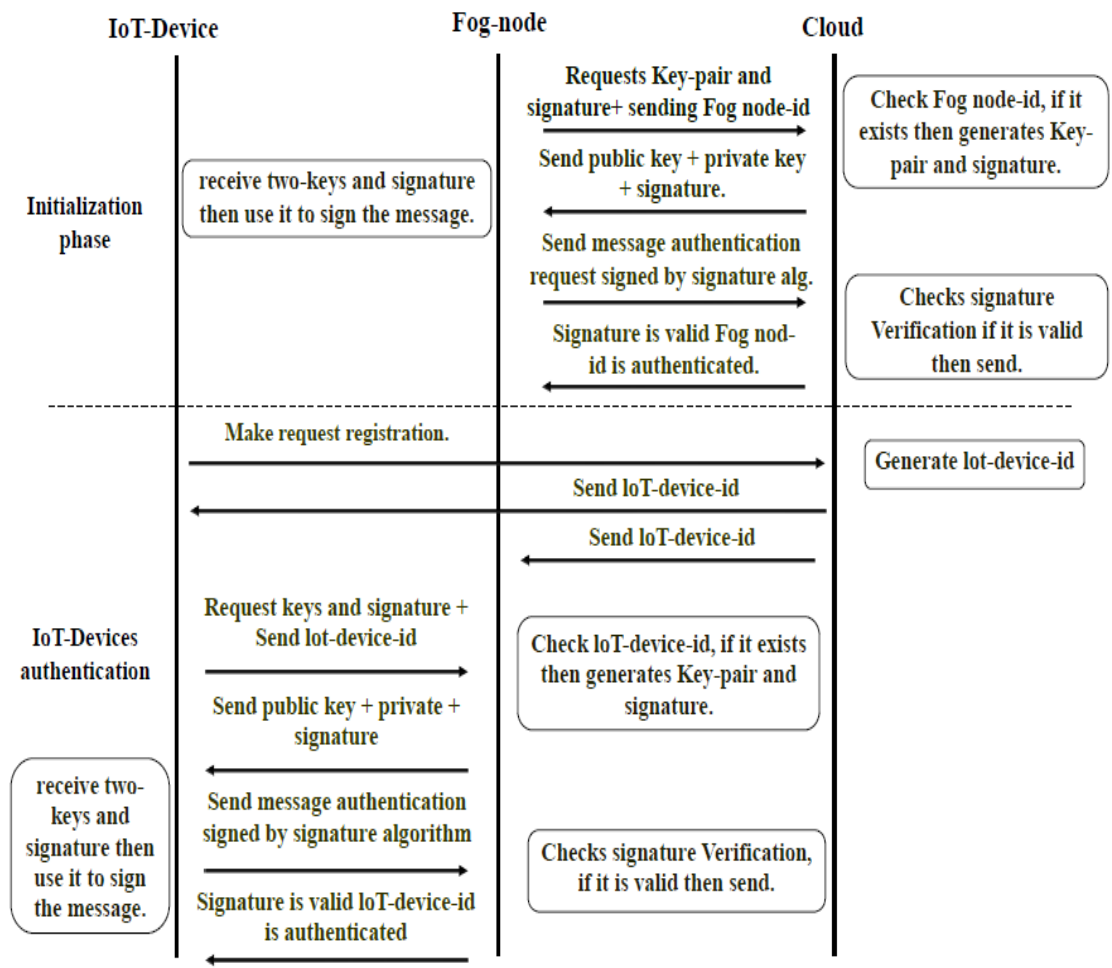

Figure 3. ECDSA Authentication Scheme.

International Journal of Computer Networks & Communications (IJCNC) Vol.15, No.4, July 2023

## 5. RESULTS IMPLEMENTATION AND ANALYSIS

In this section, we perform several experiments to study the performance of the proposed ECDSA authentication scheme under different scenarios. First, we introduce the experiment and the results of executing the five types of ECDSA. Then, the simulation results using OMNET++ [34] for the five types of ECDSA are presented and analysed.

In this paper, we have performed several experiments to study the performance of the proposed ECDSA authentication scheme. We have implemented our protocol ECDSA using laptop-device with specifications (core i7 $7^{th}$ and ram 8 GB) and analysed the performance of five curve types used in authenticating the IoT-devices. The performance is measured in terms of computation time and energy consumption. In these experiments, we have implemented our protocol ECDSA using different numbers of IoT-devices ranged from 20 to 500. Finally, the simulation compared the computation time and energy consumption of our proposal using different numbers of IoT-devices with RSA and Lindell protocols.

The performance of lightweight authentication using the ECDSA scheme is evaluated using various measures such as key size, energy consumption, and the time of key generation functions, hash functions, signature generation, and verification. To analyse the curves, ECDSA is utilised for the selected curves to compute the time and energy consumption required to perform all the ECDSA operations.

Table 2. Key size of five curve types implementation.

| Curve | M-221 | Secp256r1 | Curve25519 | Brainpool P256t1 | M-511 |
|---|---|---|---|---|---|
| Key size | 221-bit | 256-bit | 255-bit | 256-bit | 511-bit |

As we see in Table 2, we show the key size of the five curve types that we use in applying ECDSA on each curve.

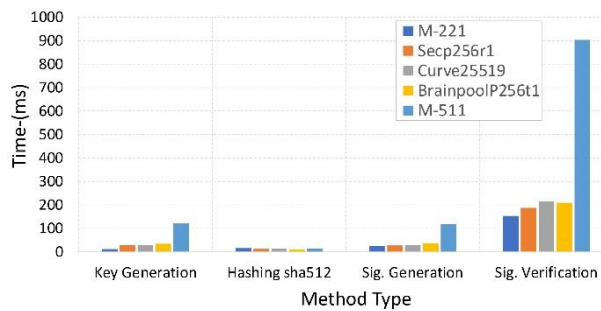

Figure 4. Time taken by each curve for ECDSA.

We can see in Figure 4, the results of implementing our protocol ECDSA using laptop-device with specification (core i7 7th and ram 8GB). The average value of the computation time in milliseconds for the implementation of five curves of ECDSA, which consists of key generation function time, SHA-512 function time, signature generation, and signature verification functions time. We see that in the case of curve M-221, it is less due to the small key size and use Montgomery elliptic curve standard, which is better than the other standard curves. On the other hand, the curve M-511 uses a large amount of time compared to any curve we use due to the largest key size. So, the M-221 curve is suitable for IoT applications.

International Journal of Computer Networks & Communications (IJCNC) Vol.15, No.4, July 2023

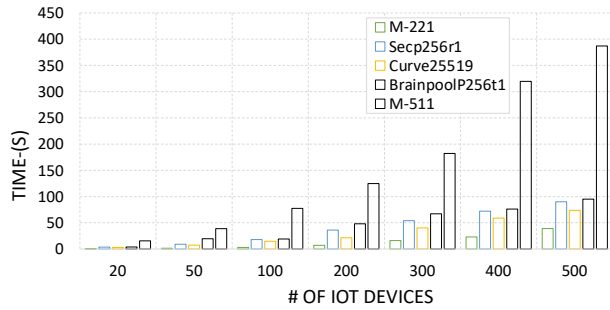

Figure 5. Computation Time of ECDSA curve types with different # of IoT-devices.

Figure 5 shows the computation time of ECDSA curve types with a different number of IoT-devices. The computation time is the average value of the computation time in seconds for the implementation of the five curve types of ECDSA. From this figure, we can observe that M-221 curve computations times are small for a different number of IoT-devices compared with the other curves due to its small key size and curve standard type.

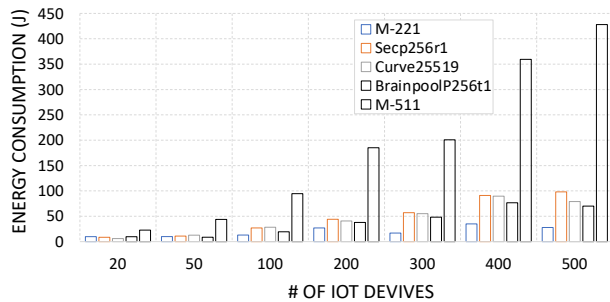

Figure 6. Energy consumption of ECDSA curve types with different # of IoT-devices

Figure 6 compares the energy consumption of ECDSA curve types and a different number of IoT-devices, the Energy consumption is the total amount of energy required for the implementations of ECDSA phases and is measured in joule (J). From this figure, we note that the M-221 curve consumes a small amount of energy compared with the other curves. This is because the M-221 curve takes a small-time to generate and verify its signature than the other curve types.

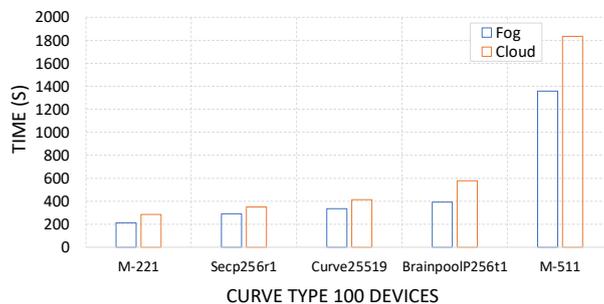

Figure 7. Computation time of ECDSA curve types running in Fog and Cloud.

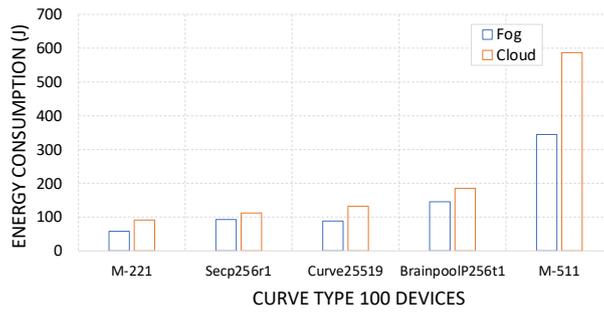

Figure 8. Energy consumption of ECDSA curve types running in Fog and Cloud.

Figure 7 and Figure 8, show the computation time and energy consumption of ECDSA curve types running in Fog and Cloud using 100 IoT-devices. The computation time in seconds and energy consumption in KJ. It can be seen that Fog requires less time and energy than Cloud due to the advantages of Fog, such as low latency, no problems with bandwidth, loss of connection is impossible and power efficiency.

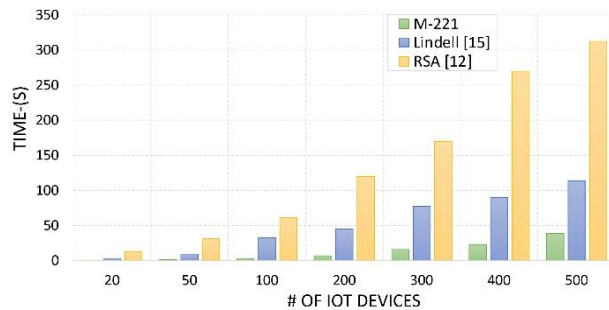

Figure 9. Computation time of RSA [12], Lindell [15] and our proposal(ECDSA M-221 curve).

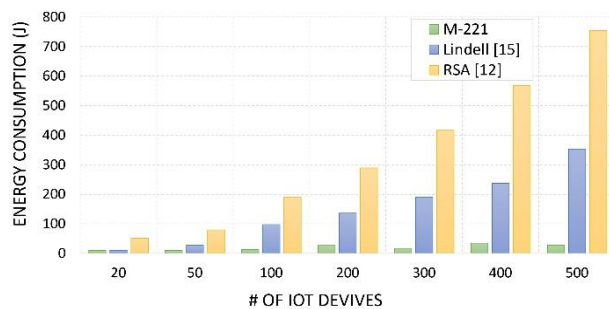

Figure 10. Energy consumption of RSA [12], Lindell [15] and our proposal (ECDSA M-221 curve).

Figure 9 and Figure 10, show the computation time in seconds and the amount of energy consumption in joule (J) of implementing each of RSA [12], Lindell [15], and our proposal (ECDSA M-221 curve) as a function of the different number of IoT-devices. It can be seen that when using 20 IoT-devices, the time and amount of energy are relatively close, but when the number of devices increases, there is a massive difference in time and energy due to the requirements of RSA, which depends on the integer factorization problem. So, the RSA needs large cryptography keys for strong security of data, which require complex calculations that are considered to be too computationally expensive for memory-constrained devices or small devices. And Lindell [15] use artificial and interactive assumption on Paillier scheme and sending encrypted secret share with Paillier requires additional highly cost range checks so it take a long time and a large amount of energy in key generation phase and signature generation phase, which is not appropriate for modern communication. In this context, the ECDSA is considered suitable for IoT-devices as it uses small key sizes, which decreases computation time by 65% and 87%, and reduces energy consumption by 77% and 82% respectively, and makes it more efficient and performs faster than other protocols.

## 6. CONCLUSION

This paper studies the problem of IoT-device authentication in fog environments. In this paper, we have studied a secure and lightweight authentication scheme that is based on the ECC to authenticate the IoT-devices and the fog gateway. The scheme uses ECDSA to enhance the security of IoT-devices. ECDSA security is based on the intractability of the Elliptic Curve Discrete Logarithm Problem (ECDLP), which allows one to use much smaller groups. Smaller group sizes directly translate into shorter signatures, which is a crucial feature in settings where communication bandwidth is limited or data transfer consumes a large amount of energy. In this paper, we have chosen the safe curve types of elliptic-curve cryptography (ECC) such as M-221, SECP256r1, curve 25519, Brainpool P256t1, and M-551. These types of curves are the most secure of the other curves of ECC, as their security is based on the complexity of the ECDLP of the curve. And these types of curves exceed the complexity of the ECDLP. We have evaluated the efficiency of five secure curves of lightweight ECC using ECDSA with IoT-devices in a fog environment. From the extensive simulation, we concluded that the M-221 curve is an efficient curve that utilises less time and energy compared with other safe curves. Also, we compared our proposal (ECDSA M-221 curve) with two protocols, RSA and Lindell. The experiments verify that the M-221 curve outperforms the RSA and Lindell protocols under different experiment settings. In comparison to Lindell and RSA, the proposed ECDSA decreases computation time by 65% and 87% and reduces energy consumption by 77% and 82%, respectively. While the ECC protocol is much more efficient than RSA for signature generation and decryption. However, most ECDSA implementations require a secure random generator - if the same random value is reused (for different plaintext) then the private key parameter can simply be calculated.

In future work, the authors plan to make modification in ECDSA protocol to improve its signature scheme by reducing the time of signature generation and signature verification. This improvement will preserve the remorse's of IoT-Fog networks.

**CONFLICTS OF INTEREST**

The authors declare no conflict of interest.